\documentclass[journal]{IEEEtran}
\usepackage{tikz}
\usepackage{lettrine}
\usetikzlibrary{arrows,shapes,positioning,shadows,trees}
\tikzset{
  basic/.style  = {draw, text width=1.7cm, font=\sffamily, rectangle},
  root/.style   = {basic, rounded corners=2pt, thin, align=center,
                   fill=gray!15},
  level 2/.style = {basic, rounded corners=6pt, thin,align=center, fill=gray!3,
                   text width=7em},
  level 3/.style = {basic, thin, align=center, fill=white!60, text width=6em}
}
\usepackage{pgfplots}
\pgfplotsset{compat=1.11}
\pgfplotsset{grid style={dashed,gray}}
\usepackage{pgf-pie}
\usepackage{newtxmath}
\usepackage{filecontents} 
\usepackage{tikzscale}
\usepackage{array}
\usepackage{rotating}
\newcolumntype{P}[1]{>{\centering\arraybackslash}p{#1}}
\usepackage{blindtext}
\usepackage{graphicx}
\usepackage{amsmath}
\usepackage[linesnumbered,ruled]{algorithm2e}
\usepackage{amssymb}
\SetKwRepeat{Do}{do}{while}%
\usepackage{siunitx}
\usepackage{textcomp}
\usepackage{subfigure}
\usepackage{courier}
\usepackage{flexisym}
\usepackage{enumitem}
\usepackage{pifont}
\def\firstcircle{(0,0) circle (1.5cm)}
\def\secondcircle{(0:2cm) circle (1.5cm)}
\colorlet{circle edge}{black!50}
\colorlet{circle area}{gray!20}
\tikzset{filled/.style={fill=circle area, draw=circle edge, thick},
    outline/.style={draw=circle edge, thick}}

\begin{document}

\title{Smart Routing: Towards Proactive Fault-Handling in Software-Defined Networks}
\author{Ali~Malik,
        Benjamin~Aziz,
        Mo~Adda
        and~Chih-Heng~Ke

  \thanks{Ali Malik, Benjamin Aziz and Mo Adda are with the University of Portsmouth, School of Computing, Buckingham Building, Lion Terrace, Portsmouth PO1 3HE, United Kingdom, e-mail: \{ali.al-bdairi; benjamin.aziz; mo.adda\}@port.ac.uk} 
   \thanks{Chih-Heng Ke is with the Department of Computer Science and Information Engineering, National Quemoy University, Taiwan, e-mail: smallko@gmail.com}
   
   }
   
   \markboth{}%
{Shell \MakeLowercase{\textit{et al.}}: Bare Demo of IEEEtran.cls for Journals}

\maketitle


\begin{abstract}
Software-defined networking offers numerous benefits against the legacy networking systems through simplifying the process of network management and reducing the cost of network configuration. Currently, the management of failures in the data plane is limited to two mechanisms: \textit{proactive} and \textit{reactive}. Such failure recovery techniques are activated after occurrences of failures. Therefore, packet loss is highly likely to occur as a result of service disruption and unavailability. This issue is not only related to the slow speed of recovery mechanisms, but also the delay caused by the failure detection process. In this paper, we define a new approach to the management of fault tolerance in software-defined networks where the goal is to eliminate the convergence process altogether, rather than speed up failure detection and recovery. We propose a new framework, called \textit{Smart Routing}, which works based on the forewarning signs on failures in order to compute alternative paths and isolate the risky links from the routing tables of the data plane devices. We validate our framework through a set of experiments that demonstrate how the underlying model runs.


\end{abstract}


\section{Introduction}\label{sec:Intoduction}

\lettrine[findent=2pt]{{\textbf{T}}}{ }HE concern about the Internet ossification, which is a consequence of the growing number of variety networks (e.g. Internet of Things, wireless sensor, Cloud, etc.) that serve a huge number of clients (currently estimated about 9 billion) around the globe, has led to rethink about the existing rigid network infrastructure whether it can be replaced by a programmable one \cite{Internet_Ossification}. In this context, Software-Defined Networking (SDN) has emerged as a promising solution to tackle the inflexibility of the legacy networking systems.
Unlike traditional IP networks, SDN architectures consist of two layers: A \textit{control plane} and a \textit{data plane}.
The control plane, or sometimes called the \textit{controller}, represents the network brain and maintain a global view on the network. While, 
the data plane comprises network forwarding elements, i.e. switches and routers, that constitute the network topology. All the data plane elements are dictated by the network controller and therefore the entire nodes have to disclose their status periodically toward the controller, hence the global view comes. 
So far, OpenFlow \cite{OF} is the most widely used protocol that enables the controller to govern the SDN data plane through carrying the \textit{forwarding rules} as well as to facilitate the exchanging of signals between the two planes. 
Nowadays, communication networks play a vital role in human being's life activities as it represents the backbone for most of the current modern technologies. Since networking equipment are failure prone, some aspects like availability measurements, fault management and reliability become very important. 
This paper is mainly focused on the availability attribute in terms of fault tolerance and forecasting of failure in SDNs.
Despite SDN benefits, new challenges such as recovery from failure still require investigation in order to maximise their utility \cite{SDN_Challenges2015,Challenges2016}.
This paper presents a complementary approach that minimises the percentage of service unavailability through utilising an online failure prediction mechanism. This allows the network controller to perform the necessary reconfiguration prior the failure incidents. Although a number of works on SDN fault management have been proposed, none of them has exploited the feature of SDN global view in the context of failure prediction purposes.

The rest of the paper is organised as follows. Section  \ref{sec:Previous_work} provides an overview of literature related to various SDN fault management techniques. We define the problem statement in Section \ref{sec:Problem_statement} and the novelty of our work. We then present our model and framework in Section \ref{sec:Model}. Section \ref{sec:Experimental} and \ref{sec:Discussion} present the experimental procedure, observed result and comparison. Finally, a summary of this paper is provided in Section \ref{sec:Conclusion} with some future research directions.

\section {Related Work}\label{sec:Previous_work}

Link failure issues often occur as part of everyday routine network operations. Due to their negative impact on network Quality of Service (QoS), a considerable amount of research has been conducted to analyse, characterise, evaluate and recover from the frequent issues of network link failures. Such failures can either be unintentional (i.e. \textit{unplanned}) due to various causes like human error, natural disasters, overload, software bugs or cable cuts, or intentional (i.e. \textit{planned}) caused by the process of maintenance \cite{Markopoulou2008}.
Failure recovery is a necessary requirement for networking systems to ensure the reliability and service
availability. 
Generally, failure recovery mechanisms of carrier-grade networks are categorized into two types: \textit{protection} and \textit{restoration}. In protection, which is also know as \textit{proactive},  alternative solutions are pre-planned and reserved in advance (i.e. before a failure occurs).
By contrast, in restoration, which is also called \textit{reactive},  possible solutions are not pre-planned and will be calculated dynamically when failures occur.  Both approaches have pros and cons.

For example, the authors in \cite{Protection2012} implemented an OpenFlow monitoring function for achieving a fast data plane recovery. In \cite{Protection2013}, another protection method was proposed through using the OpenFlow-based Segment Protection (OSP) scheme. The main disadvantage of these approaches is that they consume the data plane storing capability since the more flow entries (i.e rules) that need to be stored, the more storage space that needs to be used. Current OpenFlow appliances in the market are able to accommodate up to 8000 flow entries only, due to known limitations of the Ternary Content-Addressable Memory (TCAM), hence making this kind of solutions costly \cite{RoadMap2014, ComprehensiveSurvey2015}. The installation of many attributes in the OpenFlow forwarding elements could lead to the deterioration of the process of match-and-action for the data plane nodes. Moreover, there is no guarantee that the preserved backups are failure-free; the backup path might fail before the primary one. 

Following the restoration approach, the authors in \cite{Sharma2011} and \cite{CarierGrade2011} presented OpenFlow restoration methods to recover from single link failures. Experiments were conducted on small scale network topologies that did not exceed 14 nodes. In \cite{Sharma2013}, the authors demonstrated, through extensive experiments, that OpenFlow restoration is not easily attainable within a time of 50ms, especially for large-scale networks, unless using protection techniques. In the same context, some works have utilised the concept of multiple disjoint paths to be employed as a backup. For example, CORONET \cite{CORONET2012} is presented as a fault-tolerance system for SDNs, in which multiple link failures can be resolved. The ADaptive Multi-Path Computation Framework (ADMPCF) \cite{ADMPCF2015} and HiQoS \cite{HiQoS2015} for large scale OpenFlow networks were produced as traffic engineering tools that are capable of holding two or more disjoint paths to be utilised when some network events (e.g. link failure) occur. Most of the existing works do not take into account the processing time of flow entries, i.e. insert, delete and modify of rules. Although the performance of OpenFlow devices is associated with their vendors, in \cite{OFLOPS2012} the authors stated that each single flow entry insertion ranges from 0.5ms to 10ms. However, 11ms is the minimum duration required to modify a single rule, since each modification process includes both deletion (of old rules) and insertion (of new ones) \cite{Dionysus2014}.

Unlike existing works, the authors in \cite{Heydari2016} considered the problem of minimising the time of flow entries required when diverting from an affected primary path to a backup one. Although, the presented algorithms do not guarantee the shortest path from end-to-end, nonetheless, they open a new direction that is worth exploring. Within the same context, the authors in \cite{Malik2017} produced new algorithms for minimising the required time to update rules through reducing the solution search space from the source to the destination in the affected path. Similarly, in \cite{Malik2017_cliques}, an approach to divide the network topology into non-overlapping cliques has been introduced to tackle the issue of failures in a localised manner, rather than taking a global view of the network. Both \cite{Malik2017} and \cite{Malik2017_cliques} took into account the time required to compute the alternative route in order to speed up the update operation. The main issue with the last three works is that they do not guarantee a shortest path from source to destination.

In summary, the previous studies demonstrated different methods to tackle the problem of data plane recovery from link failure incidents. A more recent survey \cite{SDN_Survey2017} outlines in detail more contributions to the area of fault management in SDNs. One can conclude that protection approaches are not ideal due to the TCAM space exhaustion problem, whereas the latency issue is the major drawback of the existing restoration approaches. As a result, we believe that more research is needed in terms of achieving efficient SDN resilience, which is the main aim of this work.
\section{Problem Statement and Contributions}\label{sec:Problem_statement}
Current SDN fault tolerance mechanisms inescapably lead to a certain amount of packet loss as well as to a certain probability of service unavailability. This is due to the delay of the convergence scheme $T_C$. We define $T_C$ as the time taken by the OpenFlow controller to amend a path in response to failure scenario. Typically, the convergence time in SDNs can be defined in terms of three factors:

$\bullet$ \textit{Failure detection time} ($T_D$): This is the required time to detect a failure incident. Compared with the conventional networking systems, the centralised management and global view of an SDN eases this task by continuously monitoring network status and obtaining notifications upon failure. However, the speed of receiving a notification is sometimes associated with the nature of network design and mode of communication (i.e. in-band or out-of-band) \cite{inoutband2015,coping2010}. According to \cite{detectiontime2014}, link failure detection time ranges from tens to hundreds of milliseconds, depending on the type of commercial OpenFlow switch being used. \newline
$\bullet$ \textit{New route computation time} ($T_{SP}$): This is the spent time when network controller runs a nominated shortest path routing algorithm (e.g. Dijkstra \cite{Dijkstra1959}) to compute the backup path (usually for the reactive fault tolerance strategies). The $T_{SP}$ computation time could reach 10s of milliseconds \cite{Malik2017} according to how big the network is.\newline
$\bullet$ \textit{Flow entries update time} ($T_{Update}$): This is the required time to update the relevant switches (i.e. nodes who are involved in the affected path). Again, this factor depends on how many forwarding rules need to be updated after the failure scenario, where the amount of time for a single rule may exceed 10ms.

Accordingly, the resulting convergence time can be calculated through the following equation:
\begin{equation}\label{eq:convergence} 
    T_C = T_D + T_{SP} + \sum_{src}^{dst} T_{Update}
\end{equation}

Currently, the classical SDN fault management methods aim to tackle the failure after it occurrence, therefore, the recovery mechanism is activated after the moment of failure and hence all the previous work proposals embroiled in a certain amount of delay according to (\ref{eq:convergence}). The only way to completely overcome the three factors of (\ref{eq:convergence}) altogether is by handling the failure before it occurs. Therefore, failure prediction is required to provide awareness about the potential future incidents as well as allowing the controller to perform the reconfiguration action in purpose of overriding failures before causing damage on some paths. Although there are a number of studies that have put efforts in the area of failure prediction, none of these (except \cite{Prediction_OSPF2013}) has exploited the information that can be gained from any prediction method to eliminate network incidents (e.g. link failures). To the best of our knowledge, \cite{Prediction_OSPF2013} is the only realistic study that discussed the advantages of failure prediction through producing a risk-aware routing method for the legacy IP networks. Our work is different from theirs in that we build a framework of proactive failure management for SDNs. Our work combines the concept of the online failure prediction with risk analysis towards maximising the network service availability.

With this context in mind, we can summarise the main contributions of this paper as follows:\newline
\noindent
$\bullet$
A new network model that allows for the forecasting of link failures by predicting their characteristics in an online fashion.  This model also combines the predictive capability with the decision making process using risk analysis.

\noindent
$\bullet$
We provide an implementation of the new model in terms of a couple of fault tolerance algorithms. We use simulation techniques to test the efficiency of these algorithms. Our simulation results prove that the proposed model and algorithms improve the service availability of SDNs.

\section{The Proposed Model}\label{sec:Model}

Anticipating failures before they occur is a promising approach for further enhancement of SDN failure management techniques, i.e. the proactive and reactive, in which the controller responds to failures when they take place. The SDN proposed model for anticipating link failure events is presented in this section. We start by outlining some notations that will be used throughout this paper, as shown in Table \ref{tab:terms}.
\begin{table}
\centering
\caption{List of notations}\label{tab:terms}
\begin{tabular}{ |c||p{6.4cm}|}
 \hline
 Symbol& Description\\
 \hline
 $src$& Source router\\
 $dst$& Destination router\\
 $A$& Service availability\\
 $U$& Service unavailability\\
 $e_{ij}$& Link traversing any two arbitrary routers $i$ and $j$\\
 $Q_{ptr}$& A pointer that points to first $e_{ij}$ in the Queue \\
 ${F}$& Failed link set\\
 ${F_R}$& Failed/affected route set \\
 ${PF_L}$& Potential failed link set\\
 ${PF_R}$& Potential failed route set \\
 ${M}$& Prediction alarm message\\
 ${CO}$& Network controller\\
 ${T_\Omega}$& Threshold of failure probability\\
 $T_\omega$ & Threshold of risk \\
 ${OF}$& OpenFlow instruction\\
 ${TP}$& True positive\\
 ${FN}$& False negative\\
 ${FP}$& False positive\\
 $CC$& Cable cut per year\\
 ${SP_x}$& Any shortest path algorithm x in terms of hops\\
 \hline
\end{tabular}
\end{table}
The network topology is modelled as an \textit{undirected graph} $G = (V, E)$; where $V$ represents the finite set of vertices (i.e. routers) in $G$ that ranges over by $\{v_i, v_j, \dots, v_z\}$ where $\{i,j, \dots, z\} \subset \{1, \dots, n\}$ for $n\in\mathbb{N}$ , and $E$ represents the finite set of bidirectional edges (i.e. links) in $G$ that denoted as $\{e_{ij}\}$ where each $e_{ij} \in E$ is an edge that enables $v_i$ and $v_j$ to connect each other. Now, we define the following test operational function ($OP$) over a link, which reflects the link state whether it's working or not:

\[  OP(e_{ij})= 
    \begin{cases} 
      1 \ \ \ \text{the link is operational}\\  
      
      0 \ \ \ \text{otherwise}\\
   \end{cases}
\]\newline
\noindent
Therefore, ${F}$ can be defined as follows:\newline

${F} = \{e_{ij}~ |~ e_{ij} \in E \wedge OP(e_{ij})=0\}$\newline

\noindent
Based on $G$, we define a path $P$ as a \textit{sequence} of vertices representing routers in the network. Each path starts from a source router, $src$, and ends with a destination router, $dst$:\newline

${P}= (src,\ldots,dst)$\newline

\noindent
We define the set $Flow$ to represent all demand traffic flows that need to be serviced. Each $flow \in Flow$ is an instance of $P$, which associates with a particular traffic that are defined by unique $src$ and $dst$ pair. We consider $flow_{set}$ to be the set of all the possible paths between $src$ and $dst$ that can be derived from $G$, which is defined as follows:\newline

${flow_{set}=\{{P}~ |~ (\textit{first}({P})=src)   \wedge  (\textit{last}({P})=dst)\}}$\newline

\noindent
and the definition of \textit{first} and \textit{last} is given as functions on any general sequence $(a_1,\ldots,a_n)$:\newline

$\textit{first}((a_1,\ldots,a_n))=a_1$, $\textit{last}((a_1,\ldots,a_n))=a_n$\newline

\noindent
We also consider ${P}_{set}$ as a set that contains all the admissible paths that can be constructed from $G$, so this means that $ P \in {P}_{set}$ and therefore, $Flow \subset {P}_{set}$.
When a link failure is reported in $G$, then, we identify the affected routes as follow:\newline

${F_R} = \{flow~ |~ flow \in Flow \wedge \exists _{{v_i, v_j}} . v_i, v_j \in flow \wedge OP(v_i, v_j)=0$\}\newline

\noindent
In the same context, but this time we consider the case of when there is a link failure prediction message $m_i \in {M}$ such that ${M}$ set denoted by $\{m_i\}_{i=1}^{n}$ where each $m_i \in M$ is defined as $m_i= (\bar{e}_{ij}, t)$, where $t$ is the time when the system receives $m_i$. In this context, we define the following:\newline

${PF_L} = \{\bar{e}_{ij}~ |~ \bar{e}_{ij} \in E \wedge \exists_{m_i}. m_i=(\bar{e}_{ij}, t) \wedge m_i \in {M} $\}\newline

\noindent
to characterise the received link, which we use $\bar{e}_{ij}$ to imply that ${e}_{ij} \in {PF_L}$ is a shorthand, with state of \textit{potential to fail} and hence it does not belong to $F$. Now, we can define the \textit{potential to fail route} set as follows:\newline

    ${PF_R} = \{\bar{flow}~ |~ \bar{flow} \in Flow \wedge (\exists_{\bar{e}_{ij}} . {\bar{e}_{ij}}   \in \bar{flow}~  \wedge {\bar{e}_{ij}}  \in {PF_L})$\}\newline

\noindent
where $\bar{flow}$ is a $flow$ that has at least one $\bar{e}_{ij}$, in other words, $\bar{flow} \cap {PF_L} \neq \emptyset $.
\subsection {SDN Predictive Model}
All the previous efforts that dealt with data plane failures have succeeded in mitigating the impact of failures (e.g. reduce the downtime) rather than attempting to obviate their effect, such as the service unavailability. Network incidents that cause routing instability, i.e. flaps, and lead to significant degrading of network service availability vary \cite{Origin1999,Joint_analysis2010}, however, we are merely concerned with the type of data link failure. By relying on monitoring techniques, some failures can be predicted through failure tracking, syndrome monitoring, and error reporting \cite{Salfner2010}. Consequently, a set of conditions can be defined as a base to trigger a failure warning when at least one of the predefined conditions is satisfied, as follows:\newline 


$\textit{if} \ \Big < \textit{condition} \Big > \ \textit{then} \ \Big < \textit{warning} \ \textit{trigger} \Big >$\newline



\begin{figure}[!htbp]
\centering
\includegraphics [scale=0.49]{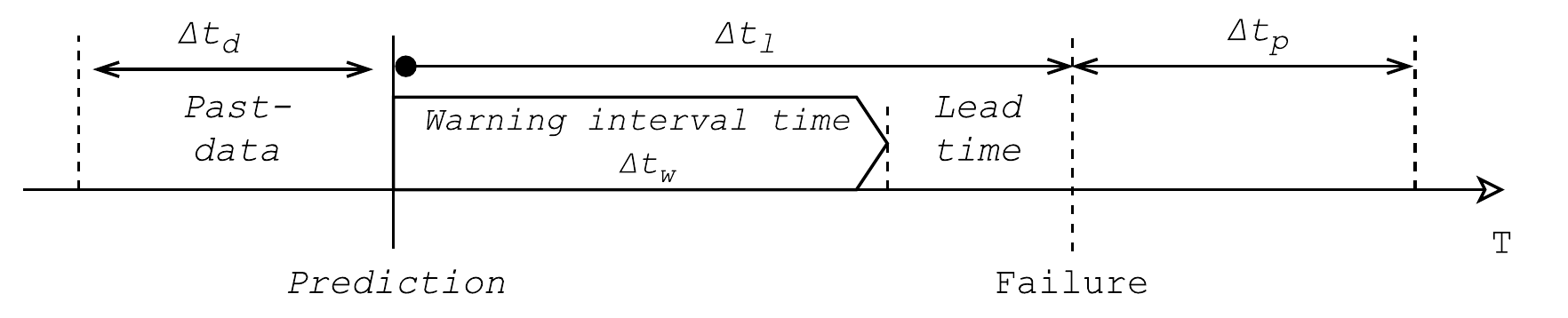}
\caption{Online failure prediction and time relations \cite{Salfner2010}.}\label{fig:prediction}
\end{figure}

\noindent
Online failure prediction strategies vary such as machine learning techniques (e.g. using the $\kappa$-nearest neighbor algorithm \cite{Prediction_machineLearning}) 
and statistical analysis methods (e.g. time series \cite{Salfner2010}, Kalman and Wiener filter \cite{kalman_wienerbook}). Such techniques can be used to predict the incoming events through relying on the past and current state information of a system. However, in this paper, we do not intend to propose a failure prediction solution as extensive studies have been conducted in this field with remarkable achievements. Instead, employing the online failure prediction as a technique to enrich the current SDN fault management is one of the main aims of this work. A generic overview of the time relations of online failure prediction is presented in Figure \ref{fig:prediction}.

\begin{itemize}
  \item $\Delta {t_d}$: represents the past (historical) data upon which the predictor is forecasting the upcoming failure events.
  \item $\Delta {t_l}$: represents the lead time upon which a failure alarm is generated. It can also be defined as the minimum duration between the prediction and failure.
  \item $\Delta {t_w}$: represents the warning time in which an action may be required to find a new solution based on the predicted event. Therefore, $\Delta {t_l}$ must be greater than $\Delta {t_w}$ so that the information from prediction will be serviceable. In SDN, the $\Delta {t_w}$ should be at least adequate to the time required to set up the longest shortest path in given $G$.
  \item $\Delta {t_p}$: represents the time for which the prediction will be assumed to be a valid case. This should be defined carefully by the network operator so as to identify the true and false alarms after a certain time window.
\end{itemize}
     The quality of the failure prediction is usually evaluated by two parameters: {\small ${FP}$} and {\small ${FN}$}; whereas, \textit{Recall} 
     and \textit{Precision} are the two well-known metrics that are used to measure the overall performance.


\begin{equation}
Recall = \frac{{TP}}{{TP} + {FN}} \ , \
    Precision = \frac{{TP}}{{TP} + {FP}}
\end{equation}

Recall is defined as the ratio of the accurately captured failures to the total number of the certainly occurred failures. However, Precision is defined as the ratio of the correctly classified failures to the total number of the positive predictions. Correspondingly, SDN controller actions will now associate with predicted and unpredicted situations as listed in Table \ref{tab:Prediction_actions}.

\begin{table}[!htpb]
\centering
\caption{Controller actions based on prediction}\label{tab:Prediction_actions}
\begin{tabular}{ |p{1.2cm}||p{4cm}|}
 \hline
 Prediction& Action\\
 \hline
 ${TP}$& Select an alternative route\\
 ${FP}$& Unnecessary/needless action\\
 ${FN}$& Call the standard failure recovery\\
 \hline
\end{tabular}
\end{table}

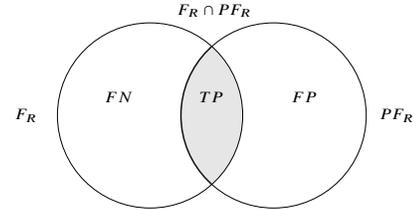
\begin{figure}[!htpb]
\centering
\scriptsize
\resizebox{.30\textwidth}{!}{%
\begin{tikzpicture}
    \begin{scope}
        \clip \firstcircle;
        \fill[filled] \secondcircle;
    \end{scope}
    \draw \firstcircle;
    \draw \secondcircle;
    \draw (-2,-0) node {${F_R}$};
    \draw (4,0) node {${PF_R}$};
    \draw (-0.5,0.3) node {${FN}$};
    \draw (2.5,0.3) node {${FP}$};
    \draw (1,0.3) node {${TP}$};
    \node[anchor=south] at (current bounding box.north) {${F_R} \cap {PF_R}$};
\end{tikzpicture}}
\caption{Relation between prediction and failure sets} 
\label{fig:4}
\end{figure}

On one hand, every false failure alarm will lead to an unnecessary reconfiguration for a particular set of routes in $Flow$ and this will cause unwitting network instability. On the other hand, a controller needs to deal with the undetected failures in a similar way to the classical methods. Consequently, the more precise behaviour of prediction, the higher the percentage of network stability and service availability will be gained. The relevance between the network model and the predictive model is summarised in Figure \ref{fig:4}.
\subsection{Failure Event Model}\label{failure_model}
We have implemented an approach of generating failure events as it is very difficult to find a public network dataset that includes some useful details like failures, hence, we adopted an alternative approach by developing our failure model. This work intends to enhance the SDN fault tolerance and resilience through maximising the network service availability. Two basic metrics have been exploited in this model: \textit{Mean Time Between Failure} ($MTBF$) and \textit{Mean Time To Recover} ($MTTR$); which are essential for calculating the availability and reliability of each network repairable component \cite{Reliability_modeling2017},\cite{Pan2003}. $MTBF$ is defined as the average time in which a particular component functions before failing, where it comes through: $\frac{\sum (start_{down\_time} - start_{up\_time})}{number \ of \ failures}$; while, $MTTR$ is the average time required to repair a failed component. Each component (i.e. link) is characterized by its own values of both  $MTBF$ and $MTTR$, which are commonly independent from other components in the network. As a consequence of lacking real data, some metrics (such as cable length and $CC$) can be alternatively used for measuring the two availability metrics. According to \cite{Pan2003}, $MTBF$ can be calculated as follows: 

\begin{equation}
\label{eq:MTBF}
    MTBF (hours) = \frac{CC \times 365 \times 24}{Cable \ Length}
\end{equation}

For instance, when $CC$ is equal to 100 km, it means that per 100 km there will be on average one cut per year. Besides this, the $MTTR$ of a link is influenced by its length \cite{Gonzalez2012}, which expresses the fact that the longer link has a higher $MTTR$ value.
On this basis, we have designed the following formula for calculating the $MTTR$ value for each link in the network.

\begin{equation}
\label{eq:MTTR}
MTTR (hours) = \gamma \times Cable Length
\end{equation}

Where $\gamma$ is defined as a parameter indicating the time required to fix the cable, which is measured by hour/kilometer format. Due to the fact that links are physically distributed in different locations and environments, therefore, $\gamma$ differs from one link to another. In other words, even if some links have the same length, their $\gamma$ could be different as it relies on the physical location and the ambient conditions. We will discuss the use of these two values in Section \ref{sec:Framework}.
\section{Risk analysis}\label{Risk_analysis}
According to \cite{kaplanRisk1981}, risk can be defined in terms of the following three questions: What scenario could occur?  what is the likelihood that scenario would occur? and what is the consequence if the scenario does occur? We next consider these questions towards formulating failure risk in SDNs.

\textit{What scenario could occur?} We define the \textit{scenario} as any undesirable event, such as failure, that breaks the service down and therefore requires a solution (e.g. path change). According to \cite{FailureScenarios2014}, there are three main types of failure scenarios, namely controller failure (including hardware and software), communication components failure (i.e. node and link) and application failure (e.g. bugs in application code), that could affect the SDN networking system. We define the set of all scenarios as $S$ ranged over by variables $s_1, s_2, \dots, s_n \in S$. 

\textit{What is the likelihood a scenario would occur?}
The likelihood that a failure scenario disrupts the network services is conditional on the occurrence of the scenario. We address this question by the aid of online failure prediction that in our case works based on a scenario's failure probability, $p\in[0,1]$.

\textit{What is the consequence if the scenario does occur?}
We address this question by computing the percentage of loss or consequence, $c$, that might potentially happen when a failure scenario is predicted at an early stage. Each failure scenario might lead to some disconnections and service disruption. Therefore, the severity of adverse effects of each failure scenario varies. For instance, $c_1$ that was caused by $s_1$ might be different from $c_2$ that was caused by $s_2$, which would reflect the outage costs that would result from disrupting some of the network connections.

Over a period of time, these questions would make a list of outcomes in the form of a triplet $\langle s_i, \ p_i, \ c_i \rangle$.
Utilising such information, $risk$ can then be formulated as a set of triples:\newline
\begin{equation}
 Risk = \{\langle s_i,\ p_i,\ c_i \rangle\}, \ \ \ i= 1, 2, \dots, n
\label{eq:generalRisk}
\end{equation}

\begin{figure}[!htpb]
\begin{center}
\includegraphics [scale=0.70]{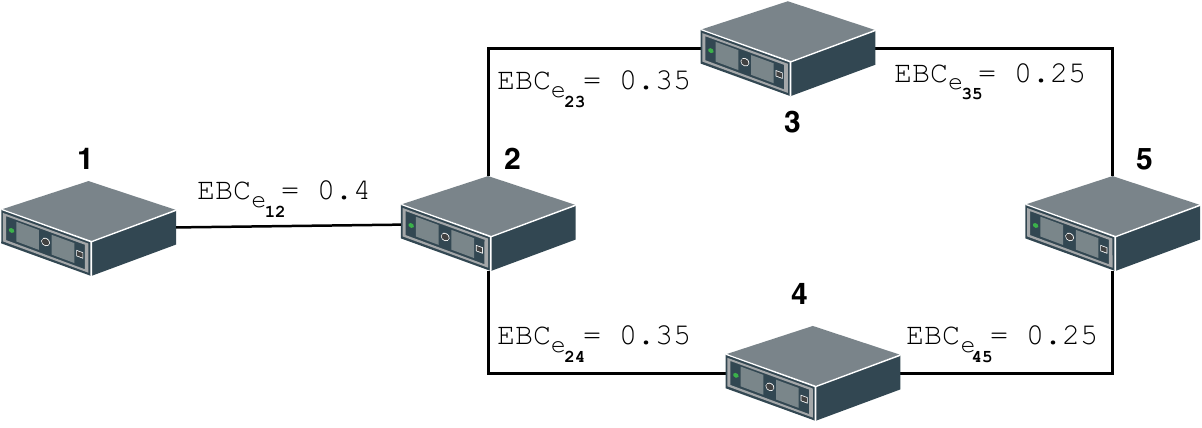}
\caption{Topology example with different EBC values\label{fig:ebc}}
\end{center}
\end{figure}

Failure scenarios may have many causes and different origins.
However, in this paper we focus only on one type, i.e. link failure scenarios that hit the data plane.
Therefore, because we are considering the only link failure scenarios, $s_{(e_{ij})}$, we shall refine the definition of risk in (\ref{eq:generalRisk}). Accordingly, we redefine risk of damage to be the combination of the probability of link failure and its consequence.
\begin{equation}\label{eq:risk}
    Risk_{s_{(e_{ij})}} = p_{(e_{ij})} \times c_{(e_{ij})}
\end{equation}
To deduce the risk value, the two factors of (\ref{eq:risk}), i.e. $p$ and $c$, can be assessed independently. On one hand, the probability, $p$, depends on the efficacy of the online failure predictor at determining the likelihood of the incoming failure scenarios, which is, in this study, defined by a selective failure probability threshold value, ${T_\Omega}$.
On the other hand $c$ can be measured based upon the percentage of affected routes that would result from the anticipated scenario. By utilising some global network topological characteristics, such as \textit{Edge Betweenness Centrality (EBC)}, the consequence score can be identified. The edge betweenness centrality of a link $e_{ij}$ is the total number of shortest paths between pairs of nodes that traverse the edge $e_{ij}$ \cite{EdgeBetweenness}, which can be formulated as follows:
\begin{equation}\label{eq:EBC}
    EBC_{e_{ij}} = \sum_{v_i \in V} \sum_{v_j \in V}\frac{\Gamma_{vi,vj} e_{ij}}{\Gamma_{vi,vj}}
\end{equation}
Where $\Gamma_{vi,vj}$ denotes the number of shortest paths between nodes $vi$ and $vj$, while, $\Gamma_{vi,vj}e_{ij}$ denotes the number of shortest paths between nodes $vi$ and $vj$ and go through $e_{ij} \in E$. For instance, Figure \ref{fig:ebc} demonstrates an example topology with an EBC value for each link in the network, which has been calculated based on Ulrik Brandes algorithm \cite{Ulrik Brandes2008}.
The network controller knows the demand traffic matrix between all pairs in the network, i.e. $Flow$. Therefore, equation (\ref{eq:EBC}) in our case is congruent with the following:
\begin{equation}
    EBC_{e_{ij\in M}}= 
    \frac{\Gamma_{flow} e_{ij}}{\Gamma_{flow}}
\end{equation}
Where $\Gamma_{flow}$ denotes the total number of paths in $Flow$ set, while, $\Gamma_{flow} e_{ij}$ denotes the number of paths in $Flow$ set and pass through $e_{ij} \in M$. 
With the above context in mind, the higher the EBC value of $e_{ij}$, which is a normalised value between $0$ and $1$, the more critical the link is and therefore, the higher the score indicating the consequences. This is because the outcome of failure for a link with high EBC will definitely lead to a huge number of path failures and therefore a higher percentage of negative impacts on the availability of network services.
Our goal in this analysis is to gauge the percentage of possible loss and provide such information to the concerned decision-making mechanism, i.e. the routing mechanism in our case. For more details about the existing risk analysis methods that fit SDNs, we refer the interested readers to  \cite{SDN_Risk2018}.
\section{Framework design}\label{sec:Framework}
\begin{figure}
\begin{center}
\includegraphics [scale=0.52]{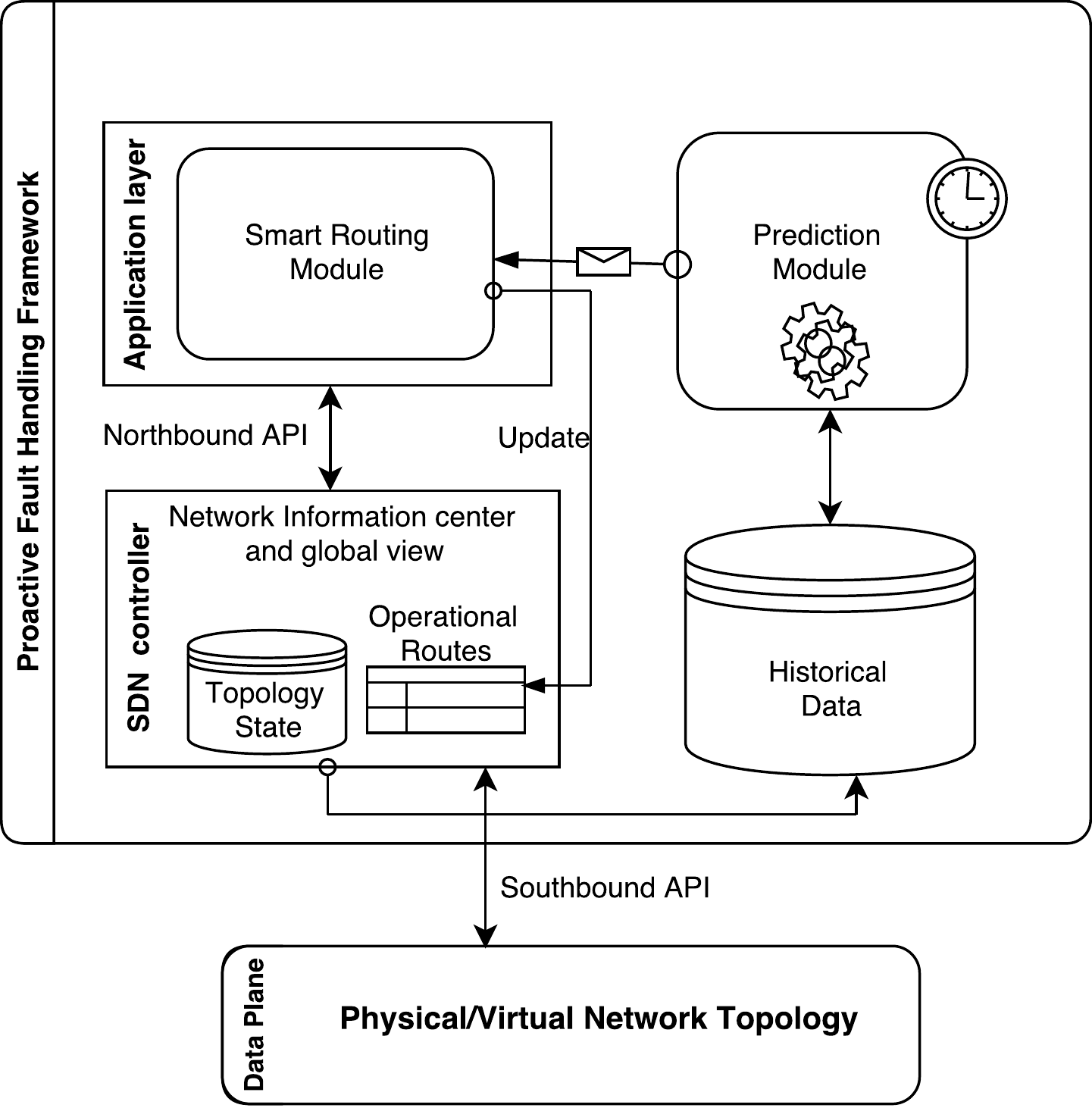}
\caption{Architecture of the proposed framework.}\label{fig:framework}
\end{center}
\end{figure}

From a high level point of view, Figure \ref{fig:framework} illustrates the main components of our proposed framework where the \textit{Smart Routing} and \textit{Prediction} modules are the primary contribution of our work. We discuss next in more detail the components we used to develop this framework.

\noindent
(a) \textit{SDN Controller} \newline
\noindent Our framework currently supports the POX controller \cite{POX_Controller}, which is an open source SDN controller written in python and it is more suitable for fast prototyping than other available controllers such as \cite{OpenFlow_Controllers2013}. The standard OpenFlow protocol is used for establishing the communication between the data and control planes, whereas the set of POX APIs can be used for developing various network control applications.

\noindent
(b) \textit{Smart Routing} \newline
Firstly, this module is responsible for maintaining and parsing the underlying network topology. Topology parameters such as the number of nodes and links, way of connection and port status can be detected via the Link Layer Discovery Protocol (LLDP) \cite{Topo_discovery2012}, which is one of the vital features of the current OpenFlow specification. The \textit{openflow.discovery}\footnote{https://github.com/att/pox/blob/master/pox/openflow/discovery.py}, which is an already developed component that can be used to send crafted LLDP messages out of OpenFlow nodes so that the topological view over the data plane layer can be constructed. This module will then convert the discovered network topology into a graph $G$ representation for efficient management purposes. To do so, we utilised the Networkx tool \cite{Networkx2008}, which is a pure python package with a set of powerful functions for manipulating network graphs. When the network starts working and after shaping the data plane topology, the shortest path for each $flow \in Flow$ is configured by the appointed ${SP_x}$ algorithm, which thereafter is stored in the \textit{Operational Routes} table that is specified to contain all the desired working (healthy) paths. In order to perceive how the link failure incident could affect the configured paths from the perspective of service availability and convergence time, we provide a simple example in Table \ref{tab:service availability} in which the service deterioration of the $flow_x$ due to link failure incident is highlighted.
\begin{table*}
\centering
\caption{Service availability and network flows relation}\label{tab:service availability}
\begin{tabular}{ |c c c c c c|}
 \hline
 Event & Flow & $src \rightarrow dst$ accessibility & $T_C$ & Serviceability & Notes\\
 \hline
 -- & $flow_x$ & Yes & -- & \ding{52} &Path is working\\
 $flow_x \in {F_R}$ & $flow_x$ & No & $T_D$ & \ding{54} & Path is not working\\
 $flow_x \in {F_R}$ & $flow_x$ & No & $T_{SP}$ & \ding{54} & Search for alternatives\\
 $flow_x \in {F_R}$ & $flow_x$ & No & $T_{Update}$ & \ding{54} & Path is restoring\\
  -- & $flow_x$ & Yes & -- & \ding{52} & Path is restored\\
 
 \hline
\end{tabular}
\end{table*}
In order to maintain the \textit{Operational Routes} table, two algorithms have been implemented each with its own view in respect to keep the $Flow$ maintained.

Algorithm \ref{alg:shortestpath} depicts the default shortest path routing strategy that is performed by the network controller. We specify Dijkstra's algorithm \cite{Dijkstra1959}, with complexity $O( |V|  + |E|  \ log \ |V|)$, as the shortest path finder approach for Algorithm \ref{alg:shortestpath}, which we denote by $SP_{\scriptscriptstyle{D}}$ instead of $SP_x$. So, the $SP_{\scriptscriptstyle{D}}$ is a Dijkstra function that can be applied on any $flow_{set}$ to return only one unique shortest path.
When the OpenFlow controller reports a link failure event, every path suffering from that failure will be detected and then two operations will be issued by the controller. First, a \textit{Remove}, denoted by ${OF}_{\scriptscriptstyle{Remove}}$, command is sent to all the routers that belong to each failed path in $Flow$ as a step to remove the incorrectly working entries, then an alternative route will be computed for every affected $flow$. The new flow entries of the alternative path are then forwarded to the relevant routers of each $flow$ through the \textit{Install}, denoted by ${OF}_{\scriptscriptstyle{Install}}$, command. Each modified $flow$, i.e. assigned to alternative, will be stored in a special set that is called \textit{the Labeled Flow} ($LF$), where: $LF \subset Flow$ and with length of $n$. This is to indicate that each $flow \in LF$ is in a sub-optimal state. The recovery from link failure procedure is demonstrated in line (1-13).
However, the algorithm also includes the reversion procedure that is activated after a failure recurs (line 15-32) and it is no less important than the recovery process \cite{Malik2018}. This procedure is required to take into account the percentage of routing flaps that is necessary for the experimental analysis. In fact, we developed this algorithm for comparison purposes only against Algorithm \ref{alg:Smart_Routing}. Therefore, it does not reflect a contribution of this paper. 
\begin{algorithm}\label{alg:shortestpath}
\scriptsize
\SetKwInOut{Input}{On Normal}
    \SetKwInOut{Output}{On Failure}

    \Input{$\forall \ {flow \in Flow} : 
    \textit{Set Primary Path as flow .} \ 
    flow \in SP_{\scriptscriptstyle{D}}(flow_{set})$}
    \Output{$\textit{Do the following procedure}$}
    
    \If {Link failure reported}
    {
    \ForEach{$e_{ij} \in {F}$}{
         \texttt{Compute:} ${F_R}$
    }
    \Do{${F_R} \neq \emptyset$}{
    
    ${OF}_{\scriptscriptstyle{Remove}} \ (flow)$
    
      ${flow}{_{set}} := {flow}{_{set}}  -  \{flow\}$
    
      $flow := SP_{D}({flow}{_{set}})$
    
    ${OF}_{\scriptscriptstyle{Install}} \ (flow)$
    
    $ LF \leftarrow flow$
    
    $F_R := {F_R}  -  \{flow\}$

    }
    }
    c := 0
    
    \If {Link repair reported}
    {
     \Do{$c \leqslant LF_{len}$}{
     \If{$flow_c$ is currently optimal}
     { 
     Do nothing
     
     c := c + 1
     }
     \If{$flow_c$ is currently sub-optimal}
     {
      ${OF}_{\scriptscriptstyle{Remove}} \ (flow_c)$
      
      $flow_c := SP_{D}({flow}_{c_{set}})$
      
      ${OF}_{\scriptscriptstyle{Install}} \ (flow_c)$
      
      ${LF} := {LF}  -  \{flow_c\}$
      
     c := c + 1
     }
     \If{number of links = $E_{len}$}
     {${LF} := empty $}
     }
    }
    
      
    
    
    \caption{Shortest Path Routing}
\end{algorithm}

Algorithm \ref{alg:Smart_Routing} is one of the main contributions of this work that exploited the prediction information towards enhancing the service availability and the fault tolerance of SDNs. This algorithm depends on Bhandari's algorithm for finding \textit{K} edge-disjoint paths \cite{Bhandari1999}, which has been utilised as a complementary to build the smart routing strategy. We denoted Bhandari's algorithm as $SP_{\scriptscriptstyle{B}}$ in place of $SP_x$.
\begin{algorithm}\label{alg:Smart_Routing}
\scriptsize
\SetKwInOut{Input}{Input}
    \SetKwInOut{Output}{Output}
    \Input{Network topology $G(V,E)$, ${M}$}
    \Output{${PF_R} \approx \emptyset$}
     
    $\forall \ {flow \in Flow} : 
    \textit{Set Primary Path as} \ 
    flow_{b_1} \ . \ flow_{b_1} \in SP_{\scriptscriptstyle{B}}(flow_{set})$
    
   \If{${M} = \{ m\}$}
      {
       ${PF_L} \gets \bar{e}_{ij}$
       }
    
    \ForEach{$\bar{e}_{ij} \in {PF_L}$}{
         \texttt{Compute:} ${PF_R}$
		 
   }
   $EBC_{\bar{e}_{ij}} = \frac{PF_{R_{len}}}{Flow_{len}}$
   
   $Risk_{\bar{e}_{ij}} = p(\bar{e}_{ij}) \times EBC_{\bar{e}_{ij}}$
   
   \If{$Risk_{\bar{e_{ij}}} \geqslant Risk_{T_\omega}$}
   {
    \Do{${PF_R} \neq \emptyset$}{
    
     ${OF}_{\scriptscriptstyle{Install}} \ (flow_{b_2} \ . \ flow_{b_2} \in SP_{\scriptscriptstyle{B}}(flow_{set}))$
     
     ${OF}_{\scriptscriptstyle{Remove}} \ (flow_{b_1} \ . \ flow_{b_1} \in SP_{\scriptscriptstyle{B}}(flow_{set}))$
     
      }

    \texttt{Wait:} $\Delta {t_p}$
    
     \eIf{$\bar{e}_{ij} \in F$}
      {
       \texttt{Mark as:} ${TP}$
       
       $LF \leftarrow {PF_R}$
       
       }
       {
       \texttt{Mark as:} ${FP}$
       
       \Do{${PF_R} \neq \emptyset$}{
      
       ${OF}_{\scriptscriptstyle{Install}} \ (flow_{b_1} \ . \ flow_{b_1} \in SP_{\scriptscriptstyle{B}}(flow_{set}))$
       
       ${OF}_{\scriptscriptstyle{Remove}} \ (flow_{b_2} \ . \ flow_{b_2} \in SP_{\scriptscriptstyle{B}}(flow_{set}))$
       }
       
       }
       }
       ${PF_R} = \emptyset$
       
      \If{$[ \ F= (e_{ij}) \land (e_{ij} \notin {M}) \ ] \  \lor \ [ \ F= ({e_{ij}) \land (e_{ij} \in M}) \land (Risk_{\bar{e_{ij}}} < Risk_{T_\omega}) \ ] $}
      {
      \texttt{Mark as:} ${FN}$
      
      \texttt{Call Algorithm1}
      }
      \If{Link repair reported}
      {
       \texttt{Call Algorithm1}
       }
    \caption{Smart Routing}
\end{algorithm}
Thereon, we consider $SP_{\scriptscriptstyle{B}}$ as a function specified to compute two link-disjoint paths with the least total cost for any given pair of nodes (i.e. $src$ and $dst$) or $flow_{set}$. For the purpose of distinguishing between the two returned paths of $SP_{\scriptscriptstyle{B}}$, we denote the first path as $flow_{b_1}$ and the second disjoint one as $flow_{b_2}$. The time complexity of $SP_{\scriptscriptstyle{B}}$ is different from the $SP_{\scriptscriptstyle{D}}$, which is a polynomial that is equivalent to $O((K+1).|E|  + |V|  \ log \ |V|)$.

The pseudo code of Smart Routing ($SR$) is demonstrated in Algorithm \ref{alg:Smart_Routing}, in which the $flow_{b_1}$ is initially selected to represent the primary path for each $flow$ in the network. The network controller will then start listening to the prediction module, which will be discussed in the next section, for the potential of future incidents. When a new message ($m$) is received, the controller will firstly identify the potential failed list, which contains the information about link which is expected to fail in the near future as described in (line 2-4). Secondly, the route (or routes) which might be affected according to the predicted failure message will be computed as a preparatory step to replace them (lines 5-7). After identifying the routes that may possibly fail, the $EBC$ for the predicted link will be calculated as a step towards measuring the risk (lines 8-10). If the risk value is below the threshold, then the prediction information will be ignored and no action will be taken. Otherwise, the flow entries of the newly computed disjoint path from the second step will be installed through using the \textit{Install} command. This is done by adjusting the disjoint path rules with lower priority than the primary path to avoid conflict of matching and action processes.

Following this step, the forwarding rules of the risky primary paths will need to be deleted in order to use TCAM resources efficiently. This needs to be done in a similar procedure to the installation but with the \textit{Remove} command as demonstrated in (lines 11-14). After swapping the primary, $flow_{b_1}$, with the disjoint, $flow_{b_2}$, this action will be considered as the correct decision for a certain period of time (i.e. $\Delta {t_p}$) as indicated in line 15. To examine the substantiality of the changing routes decision, the link that was anticipated to get down within $\Delta {t_l}$ will be compared against the failure set $F$. 
On one hand, if the link exists then, the prediction will be marked as ${TP}$. In addition, each $flow \in PF_R$ will be labeled as sub-optimal and store in 
$LF$ 
(lines 16-18). On the other hand, if the link does not exist then, the prediction will be considered as ${FP}$. In such a case, it is necessary to reset the primary path to its initial state (i.e. optimal) as deliberated in (lines 19-25). However, in case when there is a failure that is not captured by the prediction module then, it is considered as ${FN}$ and such failures are tackled by calling Algorithm \ref{alg:shortestpath} as outlined in (line 28-30). Finally, Algorithm \ref{alg:shortestpath} will also be invoked when a failed link is repaired (lines 32-34).

\begin{figure}[!htpb]
\centering
\includegraphics [scale=0.53]{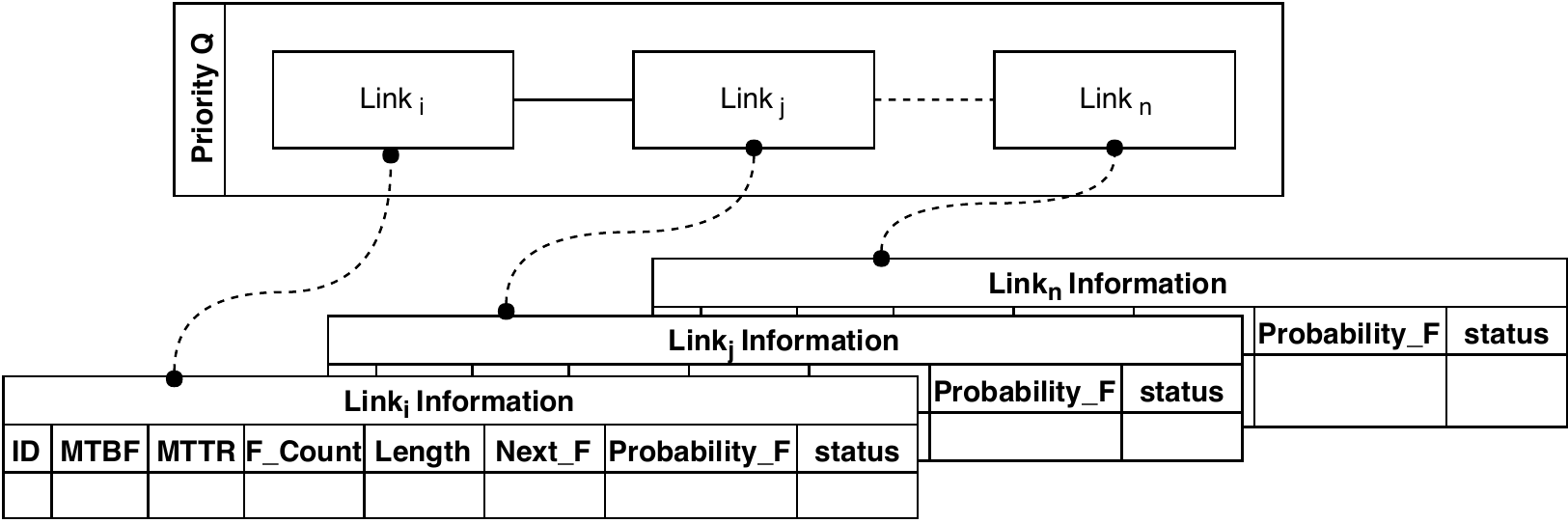}
\caption{Representation of links in priority queue}\label{fig:Queue}
\end{figure}

\noindent
(c) \textit{Prediction Module} \label{subsec:prediction_model} \newline
In this work, this module is placed on top of the parsed network topology state that gained from the network controller as a result of lacking historical data. We consider each link in the network as an independent object of link class. The link class contains a set of attributes, which currently includes eight attributes as shown in Figure \ref{fig:Queue}. 

The link attributes are used to control the up and down events. In the current implementation, we used the priority queue, $Q$, as a pool to hold all the non-faulty links. On one hand, equations (\ref{eq:MTBF}) and (\ref{eq:MTTR}) are essential for computing the two static attributes ($MTBF$ and $MTTR$) of each link. For (\ref{eq:MTBF}), we rely on the topologies information in Section \ref{subsec:topologies} and by assuming that $CC$ equals the minimum cable length in a network. While, for (\ref{eq:MTTR}) we used the uniform distribution to generate $\gamma$ for each link independently.
On the other hand, the six remaining attributes are described as follows:

\noindent
$\bullet$ ID: a numerical unique value (i.e. $1,\dots, n$) assigned to the link to represent the link identification number.\newline
$\bullet$ F\_Count: registers the number of times the link has failed.\newline
$\bullet$ Length: represents the link's length in km, which is derived from the topology specification.\newline
$\bullet$ Next\_F : refers to the \textit{next time to failure} of link, which controls the enqueue and dequeue operations of the link. In other words, this attribute determines the link's life span in the $Q$, where the link will be dequeued when \textit{Next\_F}=0.\newline
$\bullet$ Probability\_F : registers the current failure probability, $p$, of the link. For instance, the \textit{ Probability\_F} of the link ($j$) is defined as:\newline

$\frac{F\_Count(ID_j)}{\sum_{i=1}^n F\_Count(ID_i)} \times 100$\newline

\noindent
where $n$ is the $Q$ length.\newline
$\bullet$ Status : reflects the current state of the link as either operational or faulty.

On this basis, we have placed our online predictor scheme, as defined by Algorithm \ref{alg:M}, on top of the priority queue in order to send encapsulated messages about the links which satisfy the following two conditions (as described in lines 2-9): First, the probability of failure is greater than or equal to the threshold ${T_\Omega}$ and second, the leading time (i.e. $\Delta{t_l}$) is less than or equal to the \textit{next time to failure}. 

\begin{algorithm}\label{alg:M}
\scriptsize
\SetKwInOut{Input}{Input}
    \SetKwInOut{Output}{Output}
    \Input{$G(V,E)$}
    \Output{${M}$}
    \While {($Q != \emptyset$)}
       {
     \eIf{$Probability\_F_{(Q_{ptr})} \geqslant {T_\Omega}$}
      {
       \texttt{Compute:} $\Delta{t_l}$
       
       \eIf{$Next\_F_{(Q_{ptr})} \geqslant {\Delta{t_l}}$}
      {
       \texttt{Wait:} $Next\_F_{(Q_{ptr})} - \Delta{t_l}$
       
       \texttt{Generate:} $(m, \bar{e}_{ij_{(Q_{ptr})}})$
       }
       {
       $\Delta{t_l}$ is not satisfied
       }
       }
       {
       Do nothing
       }
        
          \texttt{Wait:} $Next\_F_{(Q_{ptr})} = 0$
          
       }
    
    \caption{Alarm message generator (${M}$)}
\end{algorithm}
\section{Experimental Setup and Design}\label{sec:Experimental}
Since smart routing is aimed to enhance the SDN fault tolerance in the context of network service availability, we have implemented some metrics for fair comparison between the traditional SDN and the proposed system. We also show in this section the adopted network topologies that have been utilised in our experiments.
\subsection{Availability Measurements}
Considering the convergence time that is required to shift from a failed or non-operational path to an alternative or backup one, which conforms with Equation (\ref{eq:convergence}). This convergence process definitely damages the availability of some paths, as shown in Table \ref{tab:service availability}. For the purpose of identifying the serviceable, which are denoted by "Yes", and the unserviceable, which are denoted by "No", $flows$ with respect to some failure events, we formulated this problem as follows:

\begin{center}
$(flow \cap Q) = flow \implies Yes$

$(flow \cap Q) \subset flow \implies No$
\end{center}

\noindent
where, "Yes" and "No" can be obtained by intersecting each $flow \in Flow$ against the $Q$. The $flow$ is subjected to "Yes" when all its forming edges reside in the $Q$, otherwise, the $flow$ will be considered as unserviceable and subjected to ``No". By knowing the number of serviceable and unserviceable $flows$, the service unavailability and thus the service availability can be measured. The service unavailability of SDN ($U_{SDN}$) over a given interval time with a certain number of failure events, which are denoted by $ev$, can be arrived at as follows:
\begin{equation}
    U_{SDN}(Flow, G) = \frac{\sum\limits_{\substack i=1}^{ev}{_{flow \in Flow} No}}{ev \ \times \  Flow_{len}} 
\end{equation}
\begin{footnotesize}
\end{footnotesize}

\noindent Whereas, for smart routing it is important to further consider the impact of $Recall$ values as well. Hence, the service unavailability of $SR$ ($U_{SR}$) can be arrived at through the following equation:
\begin{equation}
    U_{SR}(Flow, G) = (1-Recall) \times (U_{SDN}(Flow, G))
\end{equation}

\noindent Consequently, the availability $A_x$, with $x = \ SDN \ or \ SR$, can be arrived at through the following:
\begin{equation}
    A_x = 1 - U_x
\end{equation}

\subsection{Routing Instability Measurements}
In traditional networks, routing protocols (e.g. IGP \cite{IGP2011}) perform two routing changes as a reaction to every single failure, one time when a failure occurs and another when a failure is repaired. In fact, both changes are essential for the QoS where the first change is for the purpose of service availability, while, the goal of the second one is to return back from the backup (i.e. sub-optimal) to the primary (i.e. optimal) path again. In contrast, SDN architecture brings centralisation and programmability to the scene, therefore, traditional distributed protocols are independent of the SDN architecture. Maintaining the optimal path (e.g. minimum hops in our case) of each $flow$ will require a continuously adaptive strategy that will be responsible for replacing each sub-optimal $flow$ with the optimal one after it becomes serviceable. To do so, we assume that each alternative $flow$ is additionally stored in 
$LF$ as mentioned in Section \ref{sec:Framework}. 
For SDN, the routing flaps (denoted by $RF$) can be measured by the means of link \textit{up} (denoted by $u_f$) and \textit{down} (denoted by $d_f$) as follows:

\begin{equation}\label{RF_SDN}
    RF_{\scriptscriptstyle{SDN}} = \sum_{flow \in LF} u_f + \sum_{flow \in {F_R}} d_f 
\end{equation}
On one hand, and according to (\ref{RF_SDN}), after each link down event; a new route for each $flow \in {F_R}$ is required, which then leads to a first routing change for each $flow$. On the other hand, and after each link up announcement, the controller will need to check the state of each labeled $flow$ in $LF$ to determine if it's still the optimal choice. If so, then no change will be made, otherwise, rerouting is required and therefore it will result in another routing change.

However, for the smart routing mechanism, it is necessary to consider the three prediction parameters also (i.e. ${FN}, {TP}$ and ${FP}$) as follows:

\begin{equation}\label{RF_SR}
\footnotesize
    RF_{\scriptscriptstyle{SR}} = \sum_{flow \in {F_R}} {FN}_f + \sum_{flow \in {PF_R}} {TP}_f + \sum_{flow \in {PF_R}} {FP}_f + \sum_{flow \in LF} u_f
\end{equation}

According to (\ref{RF_SR}), the ${FN}_f$ is equivalent to $d_f$ in (\ref{RF_SDN}) as it reflects the actual failure events that have not been captured by the prediction module, while the remaining are as follows:\newline
\noindent
$\bullet$ Each true prediction will lead to a first reroute flap that gives the advantage of avoiding an upcoming failure event. While, the second flap will be similar to the scenario of $RF_{\scriptscriptstyle{SDN}}$ through inserting the $flow$ into the $LF$ and the next flap builds upon the link restoration $u_f$.\newline
\noindent
$\bullet$ Each false prediction leads into two useless flaps, one when the prediction triggers an alarm, in such a case each potential $flow$ will be added to the \textit{temporary labeled Flow} set ($TLF$), as a transient step before it recognises the prediction was false. The second flap is performed when $\Delta {t_p}$ expires.

\begin{figure}[!htpb]
\begin{center}
\includegraphics [scale=0.65]{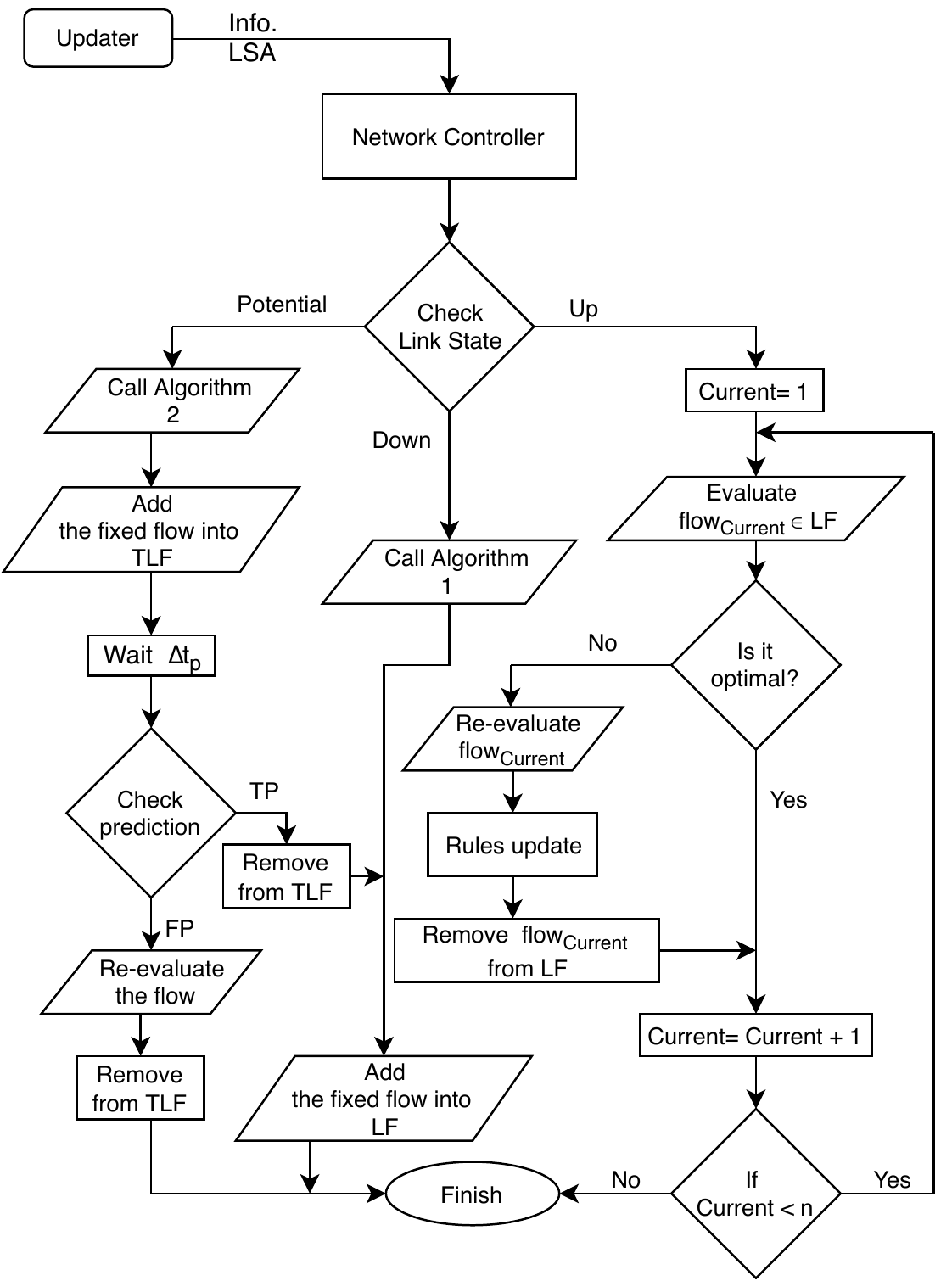}
\caption{Flow chart of routing flaps} 
\label{fig:Routing_Flaps}
\end{center}
\end{figure}

We provide an overview of the process of measuring the number of routing flaps in the flow chart of Figure \ref{fig:Routing_Flaps}, which also shows how the $LF$ is adjusted in the scenario of the two algorithms, i.e.  Algorithm \ref{alg:shortestpath} and \ref{alg:Smart_Routing}. 

Since all  actions are associated with the link state, in this work, we utilise the OpenFlow protocol to reflect the data plane links changing state by relying on the \textit{Link-State Advertisement} (LSA), in addition to the proposed prediction module that will also produce additional observed information about the potential failures. Both LSA and prediction information will be delivered to the controller through the \textit{Updater} in order to apply the appropriate action as illustrated in the flow chart.
\subsection{Simulated network topologies}\label{subsec:topologies}
In order to evaluate the proposed method, we have modelled three core network topologies as illustrated in Table \ref{tab:top}, where both janos-us and germany50 represent a real network topology instance that was defined in \cite{SNDlib}, while waxman synthetic topology
is created by the Internet topology generator Brite \cite{BRITE2001} through using the well-known Waxman model \cite{WAXMAN1988}.
\begin{table}[!htpb]
\scriptsize
\centering
\caption{Topologies' characteristics}\label{tab:top}
\begin{tabular}{| >{\centering\arraybackslash}m{0.5in} || >{\centering\arraybackslash}m{0.3in} |  |
>{\centering\arraybackslash}m{0.3in} |  |
>{\centering\arraybackslash}m{0.65in} |  |
>{\centering\arraybackslash}m{0.65in} |}
 \hline
 Topology& Nodes&  Edges& Min$_{\tiny{len}(e_{ij})}$ & Max$_{\tiny{{len}}(e_{ij})}$\\
 \hline
 janos-us& 26 &42&145 km& 1127 km\\
 germany50& 50 &88& 36 km& 236 km \\
 waxman& 70 &140& 15 km& 1099 km\\
 \hline
\end{tabular}
\end{table}

Waxman's model is a geographical approach that connects distributed routers in a plane on the basis of the distance among them, given by the following probabilistic formula:
\begin{equation}
\mathbb{P}(\{v_i,v_j\}) = \beta \ exp{^{\frac{-d(v_i,v_j)}{L \alpha}}}
\end{equation}

\noindent
where $0 < \alpha$ and $\beta \leq 1$. $d$ represents the distance between $v_i$ and $v_j$, while $L$ represents the maximum distance between any two given nodes. The number of links among the generated nodes is associated with the value of $\alpha$ in a directly proportional manner, while the edge distance increases when the value of $\beta$ is incremented. We used Brite to generate a large-scale network topology in comparison to the others (e.g. when the number of edges or nodes $\geq 100$). The characteristics of all the modelled topologies are detailed in Table \ref{tab:top}.
\subsection{Experimental Design and Implementation}
In order to validate our approach, the proposed framework is built-up on top of POX controller\footnote{The implementation code of the current framework is made available on github : \texttt{https://github.com/Ali00/SDN-Prediction-Model.}}. We evaluated our framework prototype by using the container-based emulator, Mininet \cite{Mininet2010}. Mininet is a widely used emulation system, as evidenced in a recent survey \cite{ComprehensiveSurvey2015}, for evaluating and prototyping SDN protocols and applications. It can also be used to create realistic virtual networks, running real kernel, switch and application code, on a single machine (VM, cloud or native). Our experiments were designed based on the topologies that we illustrated in the preceding section. Since one of our experimental topologies was designed via Brite, we utilised the Fast Network Simulation Setup (FNSS) \cite{FNSS2013}. FNSS is a python-based toolchain simulator that can be used to facilitate the process of network  experiments. It provides a wide range of functions and adapters that allow network researchers to parse graphs from different topology generators, such as Brite, in order to be compatible with and/or to interface with other simulator/emulator tools, such as Mininet. 

Based on the \textit{failure event model} (Section \ref{failure_model}), the general reliability theory \cite{reliabilityBook} has been utilised to generate failure events using the exponential distribution ($mean = MTBF$) for the next time to failure of each link, and lognormal distribution $E(\mu, \sigma)$ with:\newline

\noindent
$\mu = \log (MTTR) - ((0.5) \times \log(1 + ((0.6 \times MTTR)^2 / MTTR^2)))$\newline 

\noindent
and,\newline

\begin{figure}[!htpb]
\includegraphics [scale=0.73]{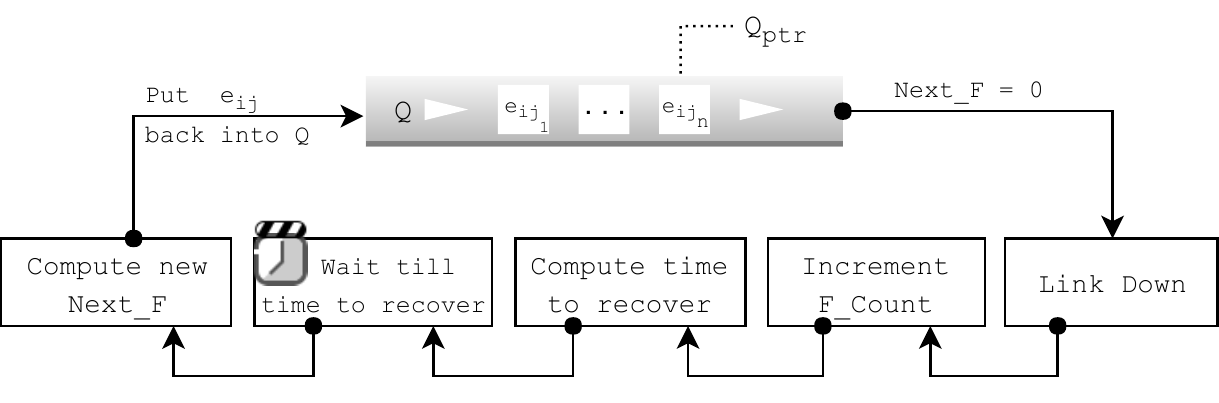}
\caption{Flow diagram of a link's life cycle in the Queue}\label{fig:flow_diagram}
\end{figure}

\noindent
$\sigma = \sqrt{\log (1+ ((0.6\times MTTR)^2/MTTR^2}$\newline

\noindent
for time to recover. Regarding failure anticipation, false and true positive have been generated during the simulated time using the uniform distribution following the specified threshold value. Figure \ref{fig:flow_diagram} summarises the simulated link queuing system that is correlated to the two metrics of reliability, i.e. MTBF and MTTR.  

In order to dispatch the prediction information that is necessarily important to the smart routing module, the distributed messages framework (ZeroMQ \cite{ZMQ}) was exploited to carry the alarm messages, ${M}$, from the prediction module to the network controller interface. In some network $flow$ conditions it will activate the smart routing module to begin a possible reconfiguration. In the emulation environment, we employed two servers; one acts as the OpenFlow controller and the other to simulate the network topologies. For each server, we used Ubuntu version 14.04 LTS running on an Intel Core-i5 processor equipped with 8 GB RAM. 
\section{Key Advantages of Smart Routing}\label{sec:Discussion}

In this section, we present comparison and evaluation of the proposed method versus the default SDN technique. To do so, the study has been conducted on the three topologies that were summarised in Table \ref{tab:top}. 
To simulate the three topologies, we ran the emulator for 144 hours, i.e.  each experimental topology was simulated in the system for 48 hours. Figure \ref{fig:results} shows the obtained results from the three topologies based on parameter settings of ${T_\Omega} = 0.25$, $T_\omega = 0.1$, $\Delta{t_l}=120s$ and $\Delta{t_p} = 30s$.
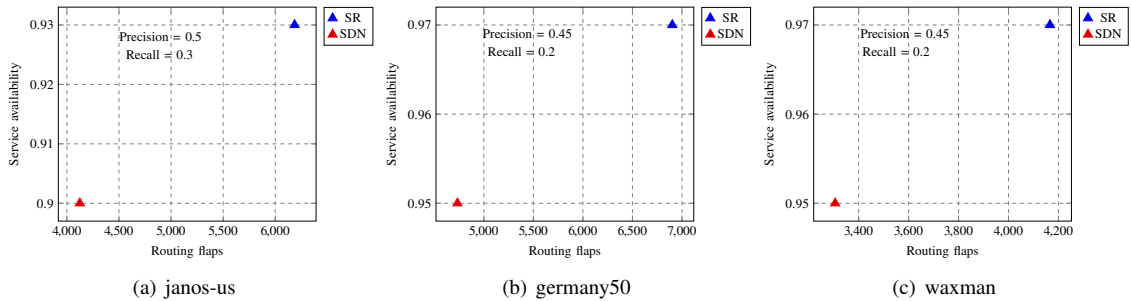
\begin{figure*}[!htpb]
\centering
\subfigure[janos-us]{
\begin{tikzpicture}[scale=0.5]
\begin{axis}[legend pos=outer north east, grid=both, xlabel={Routing flaps},
ylabel={Service availability}, mark size=4pt]
\addplot+[only marks][mark=triangle*,  scatter/use mapped color={draw opacity=0,fill=mapped color}]
coordinates {
(6186,0.93)};
\addlegendentry{SR};
\addplot+[only marks][mark=triangle*, scatter/use mapped color={draw opacity=0,fill=mapped color}]
coordinates
{(4124,0.90)};
\addlegendentry{SDN};
\node[] at (4900,0.928) {Precision = 0.5};
\node[] at (4890,0.925) {Recall = 0.3};
\end{axis}
\end{tikzpicture}}
\subfigure[germany50]{\begin{tikzpicture}[scale=0.5]
\begin{axis}[legend pos=outer north east, grid=both, xlabel={Routing flaps},
ylabel={Service availability}, mark size=4pt, ytick={0.93,0.94,0.95,0.96,0.97,0.98}]
\addplot+[only marks][mark=triangle*, scatter/use mapped color={draw opacity=0,fill=mapped color}] coordinates {
(6902,0.97)};
\addlegendentry{SR};
\addplot+[only marks][mark=triangle*,  scatter/use mapped color={draw opacity=0,fill=mapped color}]
coordinates
{(4734,0.95)};
\addlegendentry{SDN};
\node[] at (5440,0.969) {Precision = 0.45};
\node[] at (5380,0.967) {Recall = 0.2};
\end{axis}
\end{tikzpicture}}
\subfigure[waxman]{\begin{tikzpicture}[scale=0.5]
\begin{axis}[legend pos=outer north east, grid=both, xlabel={Routing flaps},
ylabel={Service availability}, mark size=4pt, ytick={0.93,0.94,0.95,0.96,0.97,0.98}]
\addplot+[only marks][mark=triangle*,  scatter/use mapped color={draw opacity=0,fill=mapped color}] coordinates {
(4166,0.97)};
\addlegendentry{SR};
\addplot+[only marks][mark=triangle*,  scatter/use mapped color={draw opacity=0,fill=mapped color}]
coordinates
{(3306,0.95)};
\addlegendentry{SDN};
\node[] at (3585,0.969) {Precision = 0.45};
\node[] at (3550,0.967) {Recall = 0.2};
\end{axis}
\end{tikzpicture}}
\caption{Routing flaps and service availability}
\label{fig:results}
\end{figure*}
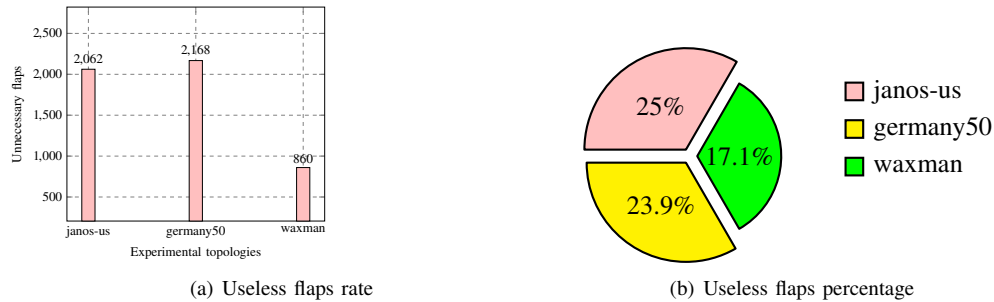
\begin{figure*}
\centering
\subfigure[Useless flaps rate]{%
\label{subfig:useless_flaps_rate}
{%
\begin{tikzpicture}[scale=0.5]
        \begin{axis}[enlarge y limits  = 0.5,
    enlarge x limits  = 0.1,
            symbolic x coords={janos-us, germany50, waxman},
            xtick=data, grid=both, nodes near coords, ylabel={Unnecessary flaps}, xlabel={Experimental topologies},
          ]
            \addplot[ybar,fill=pink] coordinates {
                (janos-us,   2062)
                (germany50,  2168)
                (waxman,   860)
            };
        \end{axis}
    \end{tikzpicture} \hspace{30mm}}}
\subfigure[Useless flaps percentage]{
\label{subfig:useless_flaps_percent}
  \begin{tikzpicture}
  \pie[polar, color={pink, yellow, green}, explode=0.1, radius =1.35, rotate=60, text=legend]
     {25/janos-us , 23.9/germany50, 17.1/waxman}
  \end{tikzpicture}
  \label{sec:SDN_trendB}
  }
\label{fig:flaps_percentage}
\caption{Routing instability measurements}
\label{fig:uselese_flaps}
\end{figure*}

As discussed earlier, the ${T_\Omega}$ and $T_\omega$ values can be selected by the network operator or by using additional algorithms (i.e. machine learning) to identify the near optimal values. 
Since the main goal of smart routing is to enhance the network service availability, we plot for each network that which gives the default SDN and SR mechanisms for the service availability percentage (Y-axis) and the rate of routing flaps (X-axis). Furthermore, for SR, the performance of the online failure predictor represented by the values of \textit{Recall} and \textit{Precision} are considered and reported respectively to each topology. In fact, \textit{Recall} value has a crucial impact on the service availability in the SR scheme, however, \textit{Precision} value has an impact on the unnecessary routing changes. It can be clearly observed that SR outperformed the default SDN in providing network service availability for all test cases. In spite of the low \textit{Recall} values (i.e. 0.2-0.3), there is still a gain in service availability. Similarly, \textit{janos-us} gained the highest improvement percentage in the service availability and this is because its \textit{Recall} value is greater than that of the other topologies.

On the other hand, the rate of the routing flaps generated by SR is always higher than the SDN. This disadvantage comes as a trade-off for improving the network service availability. Given that the routing instability by means of unnecessary flaps is correlated with the value \textit{Precision}, we have measured the only useless flaps that were generated during the simulation time and for each topology as shown in Figure \ref{fig:uselese_flaps}. Figure \ref{subfig:useless_flaps_rate} shows the only unnecessary routing changes that have been reported based on the $FP$ rate of each topology, where each single $FP$ is associated with two useless flaps, that is, one for the reconfiguration and the other for the reversion.
However, Figure \ref{subfig:useless_flaps_percent} shows the percentage of useless routing flaps for each topology in comparison with the total number of flaps. In the worst case scenario the routing flaps did not exceed 25\%. Although \textit{janos-us} topology has the highest \textit{Precision} value, it yielded a relatively high percentage of useless flaps and this is because the number of links in the topology is low, hence, it is highly likely that each single link is associated with a large number of routes in contrast to the other two topologies. It is also clearly evident that the online failure prediction plays a significant role in both service availability (by $TP$) and routing flaps (by $FP$). Based upon the experiments and simulations, we have some observations, as follows:\newline
$\bullet$ Some  alternative routes are considered as optimal after receiving an updater message, even though the received update is not involved in its conforming path. The reason for this is that the current system defines the optimal path based on the number of hops. Therefore, each alternative path that has the same number of hops as the optimal one will be considered to be an optimal path. It might not be the case if the obtained mechanism, i.e. using a specified cost function with different parameters such as bandwidth, congestion, energy, etc., is not relying on the number of hops.\newline
$\bullet$ In some cases the algorithm is barely able to find two-disjoint paths and therefore, sometimes if a path has faced two successive predictions on its links then, no change will be made. Hence, we used ($\approx$) instead of ($=$) in the output of Algorithm \ref{alg:Smart_Routing}, to imply that an entirely empty ${PF_R}$ cannot be always guaranteed.\newline
$\bullet$ It is also possible that each $flow \in LF$ may face one or more risky links, thus in such a case the entangled $flow$ state will be the same (i.e. sub-optimal).\newline
$\bullet$ In some cases and when the $Next\_F < 2 \ min$, the controller ignores the prediction if it is generated as in such a case the $\Delta {t_l}$ is not satisfied and so the controller will not have enough time for the preparation process.
\section{Conclusion and Future Work}\label{sec:Conclusion}
This paper has demonstrated how to use online failure prediction to enhance SDN service availability. 
We presented a new model for SDNs that tackles the problem of data plane link failures. Our work differs from the existing contributions by allowing SDN controllers to have a time window to reconfigure the network before the anticipated failure occurs and avoid the interruption in the availability of network services. The proposed model was implemented using a couple of new algorithms that extract the risky links from paths. Hence, when such risky links fail, no path will be affected. Our experiments were performed over a number of network topologies conducted with the link failure event model. The experimental findings demonstrate the effectiveness of the proposed method in enhancing the SDN service availability. A major drawback of this approach is the routing flaps rate that results from the failure prediction process, which may lead to network instability, especially when it reaches high rates. For this purpose, we measured the percentage of the unnecessary routing changes and in the worst scenario, it was 25\%, which we consider requires improving in future research.

For other future work, we will position the study in the setting of machine learning algorithms in order to achieve more flexibility in the decision making process, allowing this to be gauged against optimal threshold values. We are also planning to extend this work to consider disaster situations, which involve multiple link failures.

\begin{thebibliography}{1}

\bibitem{Internet_Ossification}
Lin, P., Bi, J., Hu, H., Feng, T., \& Jiang, X. (2011, November). A quick survey on selected approaches for preparing programmable networks. In \textit{Proceedings of the 7th Asian Internet Engineering Conference} (pp. 160-163). ACM.



\bibitem{OF}
McKeown, N., Anderson, T., Balakrishnan, H., Parulkar, G., Peterson, L., Rexford, J., ... \& Turner, J. (2008). OpenFlow: enabling innovation in campus networks. \textit{ACM SIGCOMM Computer Communication Review, 38}(2), 69-74.

\bibitem{Reliability_modeling2017}
Laprie, J. C. (1992). Dependability: Basic concepts and terminology. In \textit{Dependability: Basic Concepts and Terminology} (pp. 3-245). Springer, Vienna.

\bibitem{SDN_Challenges2015}
Wickboldt, J. A., De Jesus, W. P., Isolani, P. H., Both, C. B., Rochol, J., \& Granville, L. Z. (2015). Software-defined networking: management requirements and challenges. \textit{IEEE Communications Magazine, 53}(1), 278-285.

\bibitem{Challenges2016}
Akyildiz, I. F., Lee, A., Wang, P., Luo, M., \& Chou, W. (2016). Research challenges for traffic engineering in software defined networks.\textit{IEEE Network, 30}(3), 52-58.



\bibitem{Markopoulou2008}
Markopoulou, A., Iannaccone, G., Bhattacharyya, S., Chuah, C. N., Ganjali, Y., \& Diot, C. (2008). Characterization of failures in an operational IP backbone network. \textit{IEEE/ACM transactions on networking, 16}(4), 749-762.


\bibitem{RoadMap2014}
Akyildiz, I. F., Lee, A., Wang, P., Luo, M., \& Chou, W. (2014). A roadmap for traffic engineering in SDN-OpenFlow networks.\textit{ Computer Networks, 71,} 1-30.

\bibitem{Protection2012}
Kempf, J., Bellagamba, E., Kern, A., Jocha, D., Takács, A., \& Sköldström, P. (2012, June). Scalable fault management for OpenFlow. In \textit{Communications (ICC), 2012 IEEE International Conference} on (pp. 6606-6610). IEEE.

\bibitem{Protection2013}
Sgambelluri, A., Giorgetti, A., Cugini, F., Paolucci, F., \& Castoldi, P. (2013). OpenFlow-based segment protection in Ethernet networks. \textit{Journal of Optical Communications and Networking, 5}(9), 1066-1075.

\bibitem{ComprehensiveSurvey2015}
Kreutz, D., Ramos, F. M., Verissimo, P. E., Rothenberg, C. E., Azodolmolky, S., \& Uhlig, S. (2015). Software-defined networking: A comprehensive survey. \textit{Proceedings of the IEEE, 103}(1), 14-76.

\bibitem{Sharma2011}
Sharma, S., Staessens, D., Colle, D., Pickavet, M., \& Demeester, P. (2011, October). Enabling fast failure recovery in OpenFlow networks. In \textit{Design of Reliable Communication Networks (DRCN), 2011 8th International Workshop on the} (pp. 164-171). IEEE.

\bibitem{CarierGrade2011}
Staessens, D., Sharma, S., Colle, D., Pickavet, M., \& Demeester, P. (2011, October). Software defined networking: Meeting carrier grade requirements. In \textit{Local \& Metropolitan Area Networks (LANMAN), 2011 18th IEEE Workshop on} (pp. 1-6). IEEE.

\bibitem{Sharma2013}
Sharma, S., Staessens, D., Colle, D., Pickavet, M., \& Demeester, P. (2013). OpenFlow: Meeting carrier-grade recovery requirements. \textit{Computer Communications, 36}(6), 656-665.

\bibitem{CORONET2012}
Kim, H., Schlansker, M., Santos, J. R., Tourrilhes, J., Turner, Y., \& Feamster, N. (2012, October). Coronet: Fault tolerance for software defined networks. In \textit{Network Protocols (ICNP), 2012 20th IEEE International Conference on} (pp. 1-2). IEEE.

\bibitem{ADMPCF2015}
Luo, M., Zeng, Y., Li, J., \& Chou, W. (2015). An adaptive multi-path computation framework for centrally controlled networks. \textit{Computer Networks, 83,} 30-44.

\bibitem{HiQoS2015}
Jinyao, Y., Hailong, Z., Qianjun, S., Bo, L., \& Xiao, G. (2015). HiQoS: An SDN-based multipath QoS solution. \textit{China Communications, 12}(5), 123-133.

\bibitem{OFLOPS2012}
Rotsos, C., Sarrar, N., Uhlig, S., Sherwood, R., \& Moore, A. W. (2012, March). Oflops: An open framework for openflow switch evaluation. In \textit{International Conference on Passive and Active Network Measurement} (pp. 85-95). Springer Berlin Heidelberg.

\bibitem{Dionysus2014}
Jin, X., Liu, H. H., Gandhi, R., Kandula, S., Mahajan, R., Zhang, M., ... \& Wattenhofer, R. (2014, August). Dynamic scheduling of network updates. In \textit{ACM SIGCOMM Computer Communication Review} (Vol. 44, No. 4, pp. 539-550). ACM.




\bibitem{Heydari2016}
Astaneh, S. A., \& Heydari, S. S. (2016). Optimization of SDN flow operations in multi-failure restoration scenarios. \textit{IEEE Transactions on Network and Service Management, 13}(3), 421-432.

\bibitem{Malik2017}
Malik, A., Aziz, B., Adda, M., \& Ke, C. H. (2017). Optimisation methods for fast restoration of software-defined networks. \textit{IEEE Access, 5}, 16111-16123.

\bibitem{Malik2017_cliques}
Malik, A., Aziz, B., Ke, C. H., Liu, H., \& Adda, M. Virtual Topology Partitioning Towards An  Efficient Failure Recovery of Software Defined Networks. In \textit{Machine Learning and Cybernetics (ICMLC), 2017 International Conference on} (pp. 646-651). IEEE.

\bibitem{SDN_Survey2017}
Fonseca, P., \& Mota, E. (2017). A Survey on Fault Management in Software-Defined Networks. \textit{IEEE Communications Surveys \& Tutorials}.

\bibitem{inoutband2015}
Lee, S. S., Li, K. Y., Chan, K. Y., Lai, G. H., \& Chung, Y. C. (2015, October). Software-based fast failure recovery for resilient OpenFlow networks. In \textit{Reliable Networks Design and Modeling (RNDM), 2015 7th International Workshop on} (pp. 194-200). IEEE.

\bibitem{coping2010}
Desai, M., \& Nandagopal, T. (2010, January). Coping with link failures in centralized control plane architectures. In \textit{Communication Systems and Networks (COMSNETS), 2010 Second International Conference on} (pp. 1-10). IEEE.

\bibitem{detectiontime2014}
Lee, S. S., Li, K. Y., Chan, K. Y., Lai, G. H., \& Chung, Y. C. (2014, April). Path layout planning and software based fast failure detection in survivable OpenFlow networks. In \textit{Design of Reliable Communication Networks (DRCN), 2014 10th International Conference on the}(pp. 1-8). IEEE.

\bibitem{Dijkstra1959}
E. W. Dijkstra, E., W. (1959, December). A note on two problems in connexion with graphs. \textit{Numerische Mathematik, 1}(1), 269-271.

\bibitem{Prediction_OSPF2013}
Vidalenc, B., Ciavaglia, L., Noirie, L., \& Renault, E. (2013, May). Dynamic risk-aware routing for OSPF networks. In \textit{Integrated Network Management (IM 2013), 2013 IFIP/IEEE International Symposium on} (pp. 226-234). IEEE.

\bibitem{Origin1999}
Labovitz, C., Malan, G. R., \& Jahanian, F. (1999, March). Origins of Internet routing instability. In \textit{INFOCOM'99. Eighteenth Annual Joint Conference of the IEEE Computer and Communications Societies. Proceedings. IEEE} (Vol. 1, pp. 218-226). IEEE.

\bibitem{Joint_analysis2010}
Medem, A., Teixeira, R., Feamster, N., \& Meulle, M. (2010, October). Joint analysis of network incidents and intradomain routing changes. In \textit{Network and Service Management (CNSM), 2010 International Conference on} (pp. 198-205). IEEE.


\bibitem{Salfner2010}
Salfner, F., Lenk, M., \& Malek, M. (2010). A survey of online failure prediction methods. \textit{ACM Computing Surveys (CSUR), 42}(3), 10.

\bibitem{Prediction_machineLearning}
Medem, A., Teixeira, R., \& Usunier, N. (2010, December). Predicting critical intradomain routing events. In \textit{Global Telecommunications Conference (GLOBECOM 2010), 2010 IEEE }(pp. 1-5). IEEE.


\bibitem{kalman_wienerbook}
Mangoubi, R. S. (2012). \textit{Robust estimation and failure detection: A concise treatment}. Springer Science \& Business Media.

\bibitem{Pan2003}
De Maesschalck, S., Colle, D., Lievens, I., Pickavet, M., Demeester, P., Mauz, C., ... \& Derkacz, J. (2003). Pan-European optical transport networks: an availability-based comparison. \textit{Photonic Network Communications}, 5(3), 203-225.

\bibitem{Gonzalez2012}
Gonzalez, A. J., \& Helvik, B. E. (2012). Characterisation of router and link failure processes in UNINETT’s IP backbone network. \textit{International Journal of Space-Based and Situated Computing 7, 2}(1), 3-11.

\bibitem{kaplanRisk1981}
Kaplan, S., \& Garrick, B. J. (1981). On the quantitative definition of risk. \textit{Risk analysis, 1}(1), 11-27.

\bibitem{FailureScenarios2014}
Chandrasekaran, B., \& Benson, T. (2014, October). Tolerating SDN application failures with LegoSDN. In \textit{Proceedings of the 13th ACM workshop on hot topics in networks} (p. 22). ACM.

\bibitem{EdgeBetweenness}
Lu, L., \& Zhang, M. (2013). Edge betweenness centrality. In \textit{Encyclopedia of systems biology} (pp. 647-648). Springer, New York, NY.

\bibitem{Ulrik Brandes2008}
Brandes, U. (2008). On variants of shortest-path betweenness centrality and their generic computation. \textit{Social Networks, 30}(2), 136-145.

\bibitem{SDN_Risk2018}
Szwaczyk, S., Wrona, K., \& Amanowicz, M. (2018, May). Applicability of risk analysis methods to risk-aware routing in software-defined networks. In \textit{2018 International Conference on Military Communications and Information Systems (ICMCIS)} (pp. 1-7). IEEE.

\bibitem{POX_Controller}
POX Wiki. [Online]. Available at: https://openflow.stanford.edu/display/ONL /POX+Wiki.

\bibitem{OpenFlow_Controllers2013}
Shalimov, A., Zuikov, D., Zimarina, D., Pashkov, V., \& Smeliansky, R. (2013, October). Advanced study of SDN/OpenFlow controllers. In \textit{Proceedings of the 9th central \& eastern european software engineering conference in russia} (p. 1). ACM.

\bibitem{Topo_discovery2012}
Huang, W. Y., Hu, J. W., Lin, S. C., Liu, T. L., Tsai, P. W., Yang, C. S., ... \& Mambretti, J. J. (2012, March). Design and implementation of an automatic network topology discovery system for the future internet across different domains. In \textit{Advanced Information Networking and Applications Workshops (WAINA), 2012 26th International Conference on} (pp. 903-908). IEEE.

\bibitem{Networkx2008}
Schult, D. A., \& Swart, P. (2008, August). Exploring network structure, dynamics, and function using NetworkX. In \textit{Proceedings of the 7th Python in Science Conferences (SciPy 2008)} (Vol. 2008, pp. 11-16).

\bibitem{Malik2018}
Malik, A., Aziz, B., \& Adda, M. (2018, November). Towards filling the gap of routing changes in software-defined networks. In \textit{Proceedings of the Future Technologies Conference} (pp. 682-693). Springer, Cham.

\bibitem{Bhandari1999}
Bhandari, R. (1999). \textit{Survivable networks: algorithms for diverse routing}. Springer Science \& Business Media.

\bibitem{IGP2011}
Poretsky, S., Imhoff, B., \& Michielsen, K. (2011). \textit{Terminology for Benchmarking Link-State IGP Data-Plane Route Convergence} (No. RFC 6412).

\bibitem{SNDlib}
SNDlib library. [Online]. Available at: http://sndlib.zib.de.

\bibitem{BRITE2001}
Medina, A., Lakhina, A., Matta, I., \& Byers, J. (2001). BRITE: An approach to universal topology generation. In \textit{Modeling, Analysis and Simulation of Computer and Telecommunication Systems, 2001. Proceedings. Ninth International Symposium on} (pp. 346-353). IEEE.

\bibitem{WAXMAN1988}
Waxman, B. M. (1988). Routing of multipoint connections. \textit{IEEE journal on selected areas in communications, 6}(9), 1617-1622.

\bibitem{Mininet2010}
Lantz, B., Heller, B., \& McKeown, N. (2010, October). A network in a laptop: rapid prototyping for software-defined networks. In \textit{Proceedings of the 9th ACM SIGCOMM Workshop on Hot Topics in Networks} (p. 19). ACM.

\bibitem{FNSS2013}
Saino, L., Cocora, C., \& Pavlou, G. (2013, March). A toolchain for simplifying network simulation setup. In \textit{Proceedings of the 6th International ICST Conference on Simulation Tools and Techniques} (pp. 82-91). ICST (Institute for Computer Sciences, Social-Informatics and Telecommunications Engineering).

\bibitem{reliabilityBook}
Ohring, M., \& Lloyd J. R. \textit{Reliability and failure of electronic materials and devices}. Academic Press, 2009.

\bibitem{ZMQ}
ZeroMQ. [Online]. Available at: http://zeromq.org/.


\end{thebibliography}

\end{document}